\documentclass[11pt]{article}
\usepackage{epsfig}
\if@twoside\oddsidemargin25mm
\else\oddsidemargin25mm\evensidemargin25mm\marginparwidth25mm\fi
\def\thefootnote{\dag}

\def\bpl{\Big(}
\def\bpr{\Big)}
\def\ve{\varepsilon}
\def\t{\theta}

\def\w{\omega}

\def\k{\kappa}

\def\der{\partial}
\def\bq{\begin{equation}}
\def\eq{\end{equation}}
\def\brr{\begin{eqnarray}}
\def\err{\end{eqnarray}}
\def\ba{\left(\begin{array}}
\def\ea{\end{array}\right)}

\def\Dslash{\hbox{\ooalign{$\displaystyle D$\cr$\hspace{.03in}/$}}}

\newcommand{\dr}{\raise.3ex\hbox{$\stackrel{\leftarrow}{\partial }$}{}}
\newcommand{\dl}{\raise.3ex\hbox{$\stackrel{\rightarrow}{\partial}$}{}}

\newcommand{\ft}[2]{{\textstyle\frac{#1}{#2}}}

\begin{document}
\renewcommand{\a}{\alpha}
\renewcommand{\b}{\beta}
\renewcommand{\c}{\gamma}
\renewcommand{\d}{\delta}
\newcommand{\pa}{\partial}
\newcommand{\g}{\gamma}
\newcommand{\G}{\Gamma}
\newcommand{\A}{\Alpha}
\newcommand{\B}{\Beta}
\newcommand{\D}{\Delta}
\newcommand{\e}{\epsilon}
\newcommand{\E}{\Epsilon}
\newcommand{\z}{\zeta}
\newcommand{\Z}{\Zeta}
\renewcommand{\l}{\lambda}
\renewcommand{\L}{\Lambda}
\newcommand{\La}{\Lambda}
\newcommand{\m}{\mu}
\newcommand{\M}{\Mu}
\newcommand{\n}{\nu}
\newcommand{\N}{\Nu}
\newcommand{\x}{\chi}
\newcommand{\X}{\Chi}
\newcommand{\p}{\pi}
\newcommand{\R}{\Rho}
\newcommand{\s}{\sigma}
\renewcommand{\S}{\Sigma}
\newcommand{\T}{\Tau}
\newcommand{\y}{\upsilon}
\newcommand{\Y}{\upsilon}
\renewcommand{\o}{\omega}
\newcommand{\q}{\theta}
\newcommand{\h}{\eta}
\begin{titlepage}
\begin{flushright} HUB-EP 98/9 \\[1mm] 
{\tt hep-th/9801204}
\end{flushright}
\vspace{16mm}
\begin{center}
{\LARGE\bf New Consistent Limits of M-theory}                                 
\vfill
{\large Michael Faux
\footnote{faux@qft15.physik.hu-berlin.de}}\\
\vspace{7mm}
{\small
Humboldt-Universit\"at zu Berlin, Institut f\"ur Physik\\[1mm]
D-10115 Berlin, Germany\\[6pt]
 }
\end{center}
\vfill
\begin{center} {\bf Abstract} \\[3mm]
{\small
The construction of effective field 
theories describing M-theory compactified on $S^1/{\bf Z}_2$ is
revisited, and new insights into the parameters of the
theory are explained.  Particularly, the web of constraints
which follow from supersymmetry and anomaly cancelation is
argued to be more rich than previously understood.  
In contradistinction to the lore on the subject, a consistent classical
theory describing the coupling of eleven dimensional supergravity 
to super Yang-Mills theory constrained to the orbifold
fixed points is suggested to exist.}  

\end{center}
\vfill
\flushleft{January 1998}
\end{titlepage}

\renewcommand\thefootnote{\arabic{footnote}}
\setcounter{footnote}{0}

\section{Introduction}
In recent years M-theory has challenged the 
weakly-coupled heterotic string as
the most phenomenologically promising 
potentially-ultimate fundamental description of nature.
As yet, the rudiments of M-theory remain
mostly obscure.  Matrix theory and brane dynamics are two
endeavors which seem likely to bear fruit on this
issue.  Regardless of the precise microscopic
description of M-theory, of fundamental importance are the predictions
made by the theory on low-energy phenomena.
Fortunately, despite our present shortcomings in
understanding all microscopic aspects, 
we have solid insights into the low-energy regime of the theory.
It is the purpose of this paper to describe some new observations 
in this direction.

Since M-theory, appropriately compactified,
coincides with the strongly-coupled limit 
of the type IIA or the heterotic $E_8\times E_8$
string theories, there exist ample clues to the low-energy description
of the theory.  As forcefully demonstrated in \cite{wittenstring}
and \cite{hw1}, of central importance to this construction
is eleven-dimensional supergravity.
The effective theory is then further determined by
a web of constraints, involving issues from supersymmetry to
anomaly freedom, which collectively suffice to unambiguously determine at 
least certain leading terms in an effective action.  This construction has
had encouraging success in explaining away some phenomenological
shortfalls which have plagued weakly-coupled perturbative
string phenomenology.  Notable is the fact that the four-dimensional
Newton constant can take plausibly relevant magnitudes
when derived from M-theory 
\footnote{There do, however, exist other solutions to this problem,
which do not require M-theory.  See \cite{dienes} for a comprehensive
review of this issue.}.

Effective M-theories are interesting from a purely 
field-theoretical standpoint.
This is so particularly with regard to consistency.  
Exotic anomaly cancellation devices,
like the Green-Schwarz mechanism crucial for 
weakly-coupled string phenomenology, are rare.  An analogous device is
required for a viable M-theory phenomenology.
This issue was examined in \cite{hw2}, where the consistent
coupling of eleven-dimensional supergravity
to ten-dimensional super Yang-Mills theories
propagating on orbifold fixed-points 
was addressed. Central to both the weakly-coupled string effective
theories, and also to the effective M-theories
is the coupling of Chern-Simons forms to 
higher-degree tensors.  Such couplings are
generically manifest in Bianchi identities.  
For the case of weakly-coupled heterotic string theory, the 
celebrated result $dH\sim F\wedge F-R\wedge R$ is characteristic.  
The proper implementation
of an analogous relation is central to M-theory 
phenomenology.  In this paper we demonstrate a generalization
of the construction presented in \cite{hw2}, describe how this
generalization is important for the proper implementation of anomaly
cancelation, and explain how it gives rise to additional couplings 
relevant to phenomenology.

It was asserted in the original detailed work on this subject,
particularly \cite{hw2}, that there exists a consistent
coupling of eleven-dimensional supergravity to ten-dimensional 
super Yang-Mills theory propagating on orbifold fixed-points,
but {\it only if quantum effects are included}.  
The impossibilty of a consistent
classical coupling of the sort described above, has become
part of M-theory lore. In this paper we challenge that assertion, 
and suggest that a consistent classical theory may indeed exist.
This fact is directly related to the generalizations mentioned in
the previous paragraph.

In the weakly-coupled heterotic string effective theory, 
a central ingredient is the necessity for the two-form  
gauge potential in ten-dimensional $N=1$ supergravity to transform 
nontrivially under gauge transformations associated
with minimally coupled Yang-Mills supermultiplets.  
In the classical theory this
poses no obstruction to consistency because the 
simultaneous inclusion of Chern-Simons
forms, properly coupled to the tensor fields,
fully compensates for any violation of
gauge symmetry.  In the quantum theory, this feature is crucial 
for the implementation of gauge anomaly cancelation through the 
addition of counterterms not needed at the classical 
level.

In effective M-theories, it would seem plausible that
the three-form $C_{IJK}$ in eleven dimensional 
supergravity is likewise forced to transform nontrivially under
transformations $\d_\t$ associated with minimally coupled
Yang-Mills supermultiplets.
In the quantum theory this indeed turns out to be so.  
In \cite{hw2} it was assumed that even the minimal 
classical theory would require $\d_\t C_{IJK}\ne 0$.  
In that case, the $C\wedge G\wedge G$ interaction,
which is unavoidable in eleven-dimensional supergravity, 
would violate gauge invariance.  In other words,
were the above assumptions to be true, there would be an obstruction
to constructing a classical theory which 
simultaneoulsly respects both local supersymmetry
and Yang-Mills gauge invariance.
In the quantum theory the classical obstruction dissipates
because the $C\wedge G\wedge G$ interaction becomes a 
counterterm whose variation 
exactly cancels another anomalous variation arising as a loop effect.
All of this has been argued as evidence that M-theory exists only as a quantum
theory. Although this suggestion would be interesting, there is a 
loophole which needs to be properly examined.  

The minimal coupling of eleven-dimensional supergravity to
Yang-Mills supermultiplets propagating on orbifold 
fixed-points unavoidably requires a modification to a Bianchi 
identity satisfied by the four-form field strength
$G_{IJKL}$, to include a contribution from the 
Yang-Mills field strengths $F_{AB}^{\,a}$
\footnote{In this paper $I,J,K$ are eleven-dimensional world indices
taking the values $1,...,11$, while indices $A,B,C$ refer to the
ten-dimensional subset $1,...,10$.}.
This modification follows from minimally implementing supersymmetry. 
However, the modified relation does not completely 
fix the dependence of $G_{IJKL}$ on the Yang-Mills
potentials $A_A^{\,a}$.  The general solution
to the modified Bianchi identity still allows a freedom to 
manipulate features from the purely ten-dimensional components
$G_{ABCD}$ to the mixed components $G_{(11)ABC}$.  
Thus, supersymmetry alone does not completely determine $G_{IJKL}$.
However, this additional freedom {\it is} fixed by the additional 
requirements of gauge invariance and local Lorentz invariance.
This consideration may allow for a consistent classical coupling
because in the absence of quantum anomalies it does become possible
to organize the theory to contain simultaneously a nontrivially
modified Bianchi identity and a three-form $C_{IJK}$ which remains
a Yang-Mills invariant.  
The effective theory of Ho{\v r}ava and Witten \cite{hw2} does not
take into account this possible interplay between the two types of 
components of $G_{IJKL}$.

It is useful to consider the analogous situation in 
weakly-coupled heterotic
string theory by way of contrast. In that case,
the dependence of the ten-dimensional three-form field 
strength $H_{ABC}$ on the Yang-Mills gauge fields is completely fixed by
supersymmetry.  Because the Yang-Mills fields and
the three-form each have only ten-dimensional components, there is
no flexibility analogous to the case in M-theory.  But, in the case
of the weakly-coupled heterotic string one can nevertheless 
construct an effective theory
consistent even in the classical limit because supersymmetry alone does
not require the counterterms necessary for quantum anomaly cancelation.
Such terms are therefore omitted in that limit.

In addition to the considerations described above,
the effective M-theory construction also
requires modifications to the transformation rules
associated with eleven-dimensional supergravity and ten-dimensional
super Yang-Mills theory.
In particular, the supersymmetry transformation rule for the three-form
$C_{IJK}$ obtains contributions from the Yang-Mills gauge fields
$A_A^{\,a}$.  The modification in the rule for the mixed components
$\d^{\,'}_QC_{AB(11)}$ was described in \cite{hw2}.  
The consistent implementation
of the mechanism described above requires as well a modification to 
the supersymmetry transformation rule for the purely ten-dimensional
components $C_{ABC}$.   This is described in this paper.  

Upon publication of \cite{hw2}, much attention was immediately focussed on
the phenomenological consequences of the effective theory described in 
that paper.
Notable efforts have described M-theoretic modifications to
string threshold effects \cite{ns, low}, 
and possible M-theoretic explanations of supersymmety breaking
\cite{dg, noy, aq}.
But scant attention has been paid to various relevant
issues at the heart of the theory,
and no complete analysis has been paid to the problem of
consistency. For instance, of crucial importance is the cancellation 
of gravitational and mixed anomalies.  
Ho{\v r}ava and Witten \cite{hw2} have described the cancelation
of gauge anomalies in some detail, and have
strongly motivated the cancelation of gravitational 
and mixed anomalies. But a proof that the coefficients work precisely 
is lacking. We argue in this paper that a detailed analysis of gravitational
and mixed anomalies in M-theory, which is crucial to justify
effective theories used as the basis for
M-theory phenomenology, is incomplete.
We argue as well that the proper implementation of anomaly 
cancelation necessitates the new generalizations described several 
times already in this introduction.

To clarify this last point,
as described previously, one of the Green-Schwarz-like counterterms in the
low-energy description of M-theory is precisely the 
erstwhile-enigmatic $C\wedge G\wedge G$ interaction, which is 
present even in the minimal classical theory due to supersymmetry.  
Additional counter terms, crudely of the form $C\wedge R^4$ where 
$R$ is the Ricci two-form, are necessary for the cancellation of
gravitational and mixed anomalies. 
The cancellation of gravitational and mixed anomalies places
constraints on the coefficients of these terms.
Importantly, these constraints must be proven consistent
with other constraints posed by supersymmetry 
and the cancellation of pure gauge anomalies.
Consistency amongst all of these constraints requires
the new modifications which we present in this paper.

A mechanism similar to the one discused in this paper was presented
in \cite{dm} and also in \cite{lu}.  In those papers, as in this one,
an important ingredient is to include the general solution
to the modified Bianchi identity for $G_{IJKL}$, as opposed to the
specific, restricted solution used in the original work 
\cite{hw2}.  
 
When comparing expressions in this paper or expressions in most recent 
literature on M-theory to most supergravity literature, one often finds discrepancies in factors of the gravitational coupling $\k$. 
These discrepancies can be reconciled
with appropriate rescaling of fields.  To be clear, we list the
dimensionality of relevant objects in the table. 

\begin{figure}
\begin{center}
\begin{tabular}{|c||c|c||c||c|c|c||c|c|}
\hline
\multicolumn{1}{|c||}{} &
\multicolumn{2}{c||}{} &
\multicolumn{1}{c||}{} &
\multicolumn{3}{c||}{} &
\multicolumn{2}{c|}{}\\
\multicolumn{1}{|c||}{} &
\multicolumn{2}{|c||}{coupling} &
\multicolumn{1}{c||}{susy} &
\multicolumn{3}{c||}{supergravity} &
\multicolumn{2}{c|}{Yang-Mills}\\
\multicolumn{1}{|c||}{} &
\multicolumn{2}{c||}{constants} &
\multicolumn{1}{c||}{param} &
\multicolumn{3}{c||}{fields} &
\multicolumn{2}{c|}{fields}\\
\multicolumn{1}{|c||}{} &
\multicolumn{2}{c||}{} &
\multicolumn{1}{c||}{} &
\multicolumn{3}{c||}{} &
\multicolumn{2}{c|}{}\\
\hline
\hline
&\hspace{1cm}
&\hspace{1cm}
&\hspace{1cm}
&\hspace{1cm}
&\hspace{1cm}
&\hspace{1cm}
&\hspace{1cm}
&\hspace{1cm} \\
object & $\k$ & $\l$ & $\e$ & $e_I\,^a$ & $C_{IJK}$ & $\psi_I$ &
$A_A^a$ & $\chi^a$ \\
&&&&&&&& \\
\hline
&&&&&&&&\\
dimension & -9/2 & -3 & -1/2 & -1 & -3 & -1/2 & 0 & 3/2 \\
&&&&&&&&\\
\hline
\end{tabular}

\vspace{.4cm}
\parbox{5.5in}{Table:  The dimensions of objects relevant to the low-energy
description of M-theory, in units of length.  For example, the gravitational
coupling constant $\k$ has dimension $(\,{\rm length}\,)^{-9/2}$ and the
$E_8$ coupling constant $\l$ has dimension $(\,{\rm length}\,)^{-3}$.}
\end{center}
\end{figure}

With the conventions listed in the table,
both the gravitational coupling constant $\k$ and 
the Yang-Mills coupling constant $\l$  
completely factor out of the respective zeroth order (uncoupled)
actions.  See equations (\ref{action11}) and (\ref{action10})             
below to clarify this statement.  All modifications 
necessary for a consistent
coupling between these two sectors can be classified
in terms of ratios of these parameters.  Of particular
interest is the dimensionless combination $\l^6/\k^4$,
which serves as an important parameter in the theory. 
One finds that its value is determined by consistency. 

Throughout this paper we work in the so-called ``upstairs" picture,
which means that our eleven-dimensional spacetime is always assumed
to be the orbifold $R^{10}\times S^1/{\bf Z}_2$, described in more detail below.
In \cite{hw2} it is advocated that one can alternately
describe the spacetime in a ``downstairs" picture, 
as an eleven-dimensional manifold with two ten-dimensional boundaries.
We avoid this second interpretation in this paper.

We should clarify the relationship of the ten-dimensional
metric $g_{AB}$, which we use to raise and lower ten-dimensional
indices on fields constrained to orbifold fixed-points, 
to the eleven-dimensional metric $g_{IJ}$.  At the orbifold fixed-points
the elfbein $e_I\,^a$ can be taken block-diagonal and is defined to be 
\bq e_I\,^a\,|=\ba{cc} e_A\,^m & \\ & \phi \ea \,,
\label{elfbien}\eq
where the vertical bar indicates that the expression is being
evaluated at an orbifold fixed point.
Thus, we do not include the physically irrelevant 
conformal factor of $\phi^{-1/8}$ which would 
multiply $e_A\,^m$ in this decomposition
to ensure an Einstein-normalized gravitational kinetic action
in ten-dimensions
if we were performing a dimensional reduction. Since we are {\it not} performing a dimensional 
reduction, we absorb this conformal factor into our definition of 
$e_A\,^m$.  This is most natural in this context since it optimally
simplifies all expressions.  Furthermore it allows for the direct 
identification of the ten-dimensional supersymmetry parameter with the restriction to the fixed hyperplanes 
of the eleven-dimensional supersymmetry parameter. 

Spacetime is taken to be eleven-dimensional.  The eleventh dimension
is compact, and takes values on the interval $[-\pi,\pi]$,
with endpoints identified. Additionally a ${\bf Z}_2$ projection, which 
defines the orbifolding, enforces invariance under 
$x^{11}\rightarrow -x^{11}$.  There are two ten-dimensional 
hyperplanes which are fixed under this projection, defined by
$x^{11}=0$ and $x^{11}=\pi$.  As a consequence of the
projection, conditions are placed on the behavior of the 
eleven-dimensional fields at the ${\bf Z}_2$ fixed-points.  The elfbein
is constrained as indicated in (\ref{elfbien}).  
Additionally, the components
of the three-form $C_{IJK}$ which have purely ten-dimensional indices
are constrained to vanish at these points.  Thus, $C_{ABC}\,|=0$.
The components of the gravitino field $\psi_I$ 
with ten-dimensional indices are constrained to a specific
chirality from the ten-dimensional point of view.  Thus,
\bq \Gamma_{11}\psi_A\,|=\pm \psi_A\,| \,.
\eq
The remaining component $\psi_{11}$ has the opposite chirality
at the ${\bf Z}_2$ fixed points.  Half of the supersymmetry 
is broken at the fixed
points by the projection; the supersymmetry parameter satisfies
$\Gamma_{11}\e\,|=\pm\e\,|$.

This paper is organized as follows.  

In section 2 we describe a
systematic method for deriving the consistent coupling of
eleven-dimensional supergravity to ten-dimensional super Yang-Mills theory
propagating on the fixed hyperplane defined by $x^{11}=0$.  An identical
derivation would pertain to the couplings at the other fixed hyperplane
defined by $x^{11}=\pi$.  This derivation, and indeed this entire paper, is
inspired by the analogous derivation in \cite{hw2}.  But the systematics 
which we employ differ somewhat, and our results include an important
generalization to the results of that previous work.  We demonstrate
how a consistent classical coupling may exist, which does not require
the three-form $C_{IJK}$ to transform nontrivially under Yang-Mills
transformations.

In section 3 we explain how quantum effects modify the constraints 
posed by supersymmetry and by Yang-Mills gauge invariance.  
Unlike the classical case discussed in section 2, in the quantum
theory the three-form is shown to necessarily transform 
nontrivially under Yang-Mills transformations.  We indicate that the 
implementation of supersymmetry and Yang-Mills gauge invariance
is not sufficient to fix all couplings in the quantum theory.  
There remains a singe real parameter which is constrained to
be a real root of a particular cubic equation defined by
the order parameter $\l^6/\k^4$.  This freedom is fixed, however,
by requiring the absence of gravitational and mixed anomalies.  

In section 4 we describe how the results of sections 2 and 3 are
relevant to M-theory phenomenology, and we make 
some concluding remarks.

\section{The Classical Limit}
As a first step in constructing the low-energy limit of M-theory,
we study the coupling of eleven-dimensional supergravity
to ten-dimensional super Yang-Mills theory 
propagating on the fixed points of the orbifolding described in
the introduction. We first attempt to construct 
this as a purely classical theory. It is shown that such a 
coupling may indeed exist.

A systematic approach to solving this problem is to first
glean as much information about the theory and
its couplings from a study of the gauge superalgebra.  
If we were working with off-shell representations of 
supersymmetry it would be possible to derive all of the relevant
transformation rules without recourse to an action.
The construction of an invariant action would then constitute
an independent venture.  Unfortunately, given the state of the art
of supergravity theory, we are constrained to work with on-shell
representations in eleven and ten dimensions.  So the issues of 
determining the transformation rules and obtaining the invariant 
action become intimate with each other. 
Nevertheless, a useful observation is that even for on-shell theories
the gauge superalgebra closes when acting on bosonic fields.  We can
use this fact to our advantage, as we describe below.

Although, a-priori, we know the transformation rules 
and the invariant action for 
eleven-dimensional supergravity theory and similarly for the globally
supersymmetric ten-dimensional Yang-Mills theory, we do not know
the modifications which are necessary to 
describe a consistent coupling between these theories.
In effect, we need to resolve self-consistently both the 
complete superalgebra and its representation in terms
of specific transformation rules.  
For instance, a pair of supersymmetry 
transformations should commute into a particular combination of 
transformations in the full gauge algebra of the theory.  
For eleven-dimensional supergravity these
constitute general coordinate transformations $\d_{{\rm g.c.t.}}(\z)$,
supersymmetry transformations $\d_Q(\e)$,
Lorentz transformations $\d_L(\ve^{ab})$, and also tensor transformations
$\d_\Sigma(\Sigma)$ which act on the three-form as 
$\d_\Sigma(\Sigma) C_{IJK}=3\,\der_{[I}\Sigma_{JK]}$.
For the case at hand, we must include the Yang-Mills transformations
which act on the gauge fields propagating on the orbifold fixed-points, 
as $\d_\t A_A^a = D_A\t^a$.  A closed gauge algebra necessarily includes 
the following commutation relation,
\bq [\d_Q(\e_1),\d_Q(\e_2)]=\d_{{\rm g.c.t.}}(\z)
    +\d_L(\ve)    
    +\d_Q(\e_3) 
    +\d_\Sigma(\Sigma)
    +\d_\t(\t) \,,
\label{alg11}\eq
where $\z^I, \ve^{ab}, \e_3, \Sigma_{IJ}$ and $\t^a$ are 
combinations of $\e_1$ and $\e_2$ which parameterize covariant coordinate
transformations, Lorentz transformations, supersymmetry transformations, 
tensor gauge transformations and Yang-Mills gauge transformations 
respectively. 

The case of eleven-dimensional supergravity and 
the case of ten-dimensional Yang-Mills theory
represent independent solutions to the problem, with different expressions
for the dependent parameters on the right-hand-side of (\ref{alg11}).
We seek a solution which includes the complete set of fields from 
each of these theories, but which satisfies (\ref{alg11}) with
a common and unique set of parameters.  The independent solutions
just mentioned comprise the zeroth-order solution to our problem.  
So we begin by reiterating these known results.

\pagebreak

\noindent
{\it D=11 Supergravity:}\\[.1in]
One solution to (\ref{alg11}) is the case of eleven-dimensional
supergravity \cite{cjs}, which involves the elfbein $e_I\,^a$, a three-form
gauge potential $C_{IJK}$ and a spin 3/2 gravitino field $\psi_I$.
The supersymmetry transformation rules are given by
\brr \d e_I\,^a &=& \ft12\,\bar{\e}\,\Gamma^a\psi_I \nonumber\\
     \d C_{IJK} &=& -\frac{\sqrt{2}}{8}\,\bar{\e}\,\Gamma_{[IJ}\psi_{K]}
     \nonumber\\
     \d\psi_I &=& D_I(\hat{\w})\e
     +\frac{\sqrt{2}}{288}\,(\Gamma_I\,^{JKLM}-8\d_I\,^J\Gamma^{KLM})\,\e\,
     \hat{G}_{JKLM} \,,
\label{rules11}\err 
where $\hat{G}_{IJKL}$ is a supercovariant field strength given by
$\hat{G}_{IJKL}=24\der_{[I}C_{JKL]}
    +\frac{3}{\sqrt{2}}\,\bar{\psi}_{[I}\Gamma_{JK}\psi_{L]}$,
and where $\hat{\w}_I^{ab}$ is a supercovariantized spin connection.
In this solution we do not include Yang-Mills fields, and so the
transformation $\d_\t$ which appears in (\ref{alg11}) identically vanishes.
The invariant action is given by
\brr S_{SG}^{(11)} &=& \k^{-2}\,\int_{M^{11}}\,\bpl
     -\ft12\,e^{(11)}\,{\cal R}^{(11)}(e,\w)
     -\ft12 \,e^{(11)}\,\bar{\psi}_I\Gamma^{IJK}D_J(\w)\psi_K
     -\ft{1}{48}\,e^{(11)}\,G_{IJKL}^2 
     \nonumber\\[1.5mm]
     & & \hspace{.8in}
     -\frac{\sqrt{2}}{192}\,e^{(11)}\,(
     \bar{\psi}_I\Gamma^{IJKLMN}\psi_J
     +12\bar{\psi}^K\Gamma^{LM}\psi^N)G_{KLMN}
     \nonumber\\[1.5mm]
     & & \hspace{.8in}
     -\sqrt{2}\,\, C\wedge G\wedge G
     +{\rm (4\!-fermi)} \,\bpr \,.
\label{action11}\err
The four-form field strength is given by
$G_{IJKL}=24\der_{[I}C_{JKL]}$, while the supercovariant field 
strength $\hat{G}_{IJKL}$ is given above..  
In the sequel,  when we discuss modifications to $G_{IJKL}$, we refer to
the unhatted object.  The analogous hatted object can be
then be inferred from the ultimate transformation rules.   
The four-fermi terms in the action can be completely absorbed by replacing 
$\w\rightarrow \ft12(\w+\hat{\w})$ in the gravitino kinetic term, and by
replacing $G\rightarrow \ft12(G+\hat{G})$ in the $\psi^2 G$ interaction 
term.   The parameter $\k$ is the gravitational coupling constant, which
has dimensions of $({\rm length})^{-9/2}$.  We have chosen the
dimensionalities of our fields so that this parameter acts as an overall multiplicative factor.  

\vspace{.2in}
\noindent
{\it D=10 Super Yang-Mills Theory:}\\[.1in]
Another solution to (\ref{alg11}) is the case of ten-dimensional 
super Yang-Mills theory, which involves gauge potentials $A_A^a$ and 
spin 1/2 Majorana-Weyl gaugino fields $\chi^a$.
For our purposes, these are taken to transform in the adjoint
representation of the gauge group, which is ultimately
shown to be $E_8$, necessarily.
The supersymmetry transformation rules are given by
\brr \d A_A^{\,a} &=& \ft12\bar{\e}\Gamma_A\chi^{\,a} 
     \nonumber\\[1mm]
     \d\chi^{\,a} &=& -\ft14\Gamma^{AB}\e\,F_{AB}^{\,a} \,,
\label{rulesym}\err
where $F_{AB}^{\,a}$ is the field strength associated with $A_A^{\,a}$.
The invariant action is given by
\bq S_{YM}^{(10)} = \l^{-2}\,\int_{M^{10}}\,\bpl
     -\ft14\,e^{(10)}\,F_{AB}^{\,2}
     -\ft12\,e^{(10)}\,\bar{\chi}\,\Dslash(\w)\chi\,\bpr \,.
\label{action10}\eq
The parameter $\l$ is the gauge coupling constant, which has dimensions
of $({\rm length})^{-3}$. 
We have chosen the dimensionalities of our fields so that this parameter acts as an overall multiplicative factor.  

\vspace{.2in}
\noindent
{\it Couple d=11 Supergravity to (fixed-point) d=10 Super Yang-Mills:}\\[.1in]
Starting with the zeroth order theory, consisting of the transformation
rules given above and combined action $S_{SG}^{(11)}+S_{YM}^{(10)}$,
we seek to modify the transformation rules, and add appropriate 
interaction terms in order that the resulting action 
is invariant under an arbitrary local supersymmetry transformation.  
An equivalent requirement is that
the transformation rules represent a common algebra when acting on all 
fields in the combined theory.  As described above,
were we working with an off-shell representation, we could solve the problem
of obtaining transformation rules independently from the problem of
constructing an invariant action.  Since we work with an on-shell 
representation, we must be more careful.  In this case, the algebra need only 
close up to equations of motion.  However, the algebra does close on bosons.
We choose to exploit this fact to obtain as much information as possible before
considering the action.

We adopt the following philosophy.
First we solve for the most general solution to (\ref{alg11}), in terms
of transformation rules, that results in
a closed algebra when acting on the {\it bosonic} fields, 
$e_I\,^a, C_{IJK}$ and $A_A^a$.
Second, we demand an invariant action.  This second constraint
automatically ensures that the algebra closes up to
equations of motion on all fermionic fields.
Finally, we enforce that the action is invariant under Yang-Mills 
transformations.  This final requirement is nontrivial in both
the classical case and in the quantum case, but for different reasons.
In the classical case, one has to be careful that the $C\wedge G\wedge G$
interaction is inert, which requires that
$C_{IJK}$ not transform under the Yang-Mills transformations.  In the
quantum case, we must require that this term does transform, but in such 
a way as to cancel the gauge anomaly.
In the quantum case there are further constraints from 
gravitational and mixed anomalies.

\subsection{Close Algebra on Bosons}
In this subsection, we determine the set of modifications to the 
transformation rules which permit a closed gauge algebra when acting on 
bosonic fields.  A crucial part of the analysis is to
generalize to the Yang-Mills transformations to
act, not only on the fields of the Yang-Mills multiplet, but also
on certain components of the three-form $C_{IJK}$.  This curious
modification is anticipated from a related well-known result pertaining
to effective theories of the weakly-coupled heterotic string. 

The historic attempt to describe weakly-coupled heterotic string phenomenology 
involved a supergravity puzzle completely analogous to the one at hand.
In that case, the task was to couple the same ten-dimensional
super Yang-Mills theory which we are considering
to ten-dimensional $N=1$ supergravity.  
To consistently describe that coupling,
it proved necessary to generalize the Yang-Mills transformation
to act on the two-form potential in that supergravity multiplet. 
This was first described in the abelian case in \cite{chamseddine, bdrdwvn},
and later generalized to the nonabelian case in \cite{cm}.
In addition to being necessary for 
a consistent classical theory, this modification also enabled                
the cancelation all gauge and gravitational anomalies \cite{gs}.

Anticipating a similar necessity for our problem, we make as our first
anzatz a generalization of the Yang-Mills gauge transformation rules.
The analysis of \cite{hw2} describes such a modification as well.
In this paper we wish to reexplore the analysis of that paper in an
independently systematic way. We therefore modify the 
Yang-Mills transformation rule to include the following 
action on the three-form
\bq \d_\t C_{AB\,(11)}=\beta(x^{11})\,\t^a F_{AB}^a \,,
\label{crule}\eq
where $\b(x^{11})$ is an as-yet unspecified function of $x^{11}$.
The remaining components $C_{ABC}$ remain inert under the 
Yang-Mills transformations.

\pagebreak

We also alter the four-form field strength by the following 
modifications,
\brr G_{ABCD} &=& 24\der_{[A}C_{BCD]}+\gamma(x^{11})F_{[AB}^aF_{CD]}^a
     \nonumber\\[.1in]
     G_{(11)\,ABC} &=& 24\der_{[11}C_{ABC]}
     +6\,\b(x^{11})\,\w_{ABC} \,,
\label{gdef}\err
where $\g(x^{11})$ is another unspecified function of $x^{11}$,
and $\w_{ABC}$ is the Chern-Simons form associated with the
gauge potential $A_A$, which has the property that 
$\d_\t\w_{ABC}=3\der_{[A}(\t^a\,F_{BC]}^a)$.
This modification uniquely preserves the gauge invariance of 
$G_{IJKL}$.  The first expression in (\ref{gdef}) is gauge invariant 
for any choice of $\g(x^{11})$.  But the coefficient of the Chern-Simons term
in the second expression of (\ref{gdef}) is fixed by (\ref{crule}) and 
the transformation property of the Chern-Simons form.

Further generalizations are necessary in the quantum theory
due to the presence of gravitational anomalies.  These necessitate
the inclusion of factors of $R_{[AB}R_{CD]}$ and the Lorentz Chern-Simons
form into the definition of $G_{IJKL}$.  We avoid these concerns at
this point in our analysis.  We will return to this issue 
in the following section, where we discuss the consistent quantum theory.

Using the ansatz (\ref{crule}), and the definitions (\ref{gdef}),
we compute the commutator (\ref{alg11}) on all bosonic
fields $e_I\,^a, C_{IJK}$ and $A_A\,^a$ using the transformation rules
(\ref{rules11}) and (\ref{rulesym}), supplemented with 
sufficiently general modifications consistent with Lorentz covariance
and gauge covariance.
After some work, we find the unique solution to (\ref{alg11}),
which tells us the set of modifications necessary for the algebra to close.
In this way, we determine the following new terms in the 
supersymmetry transformation rules,
\brr \d_Q' C_{ABC} &=& -\ft{1}{12}\,\gamma(x^{11})
     \bar{\e}\,\Gamma_{[A}\chi^a\,F_{BC]}^{\,a} 
     \nonumber\\[.1in]
     \d_Q'\,C_{AB\,(11)} &=& \b(x^{11})\,
     \bar{\e}\Gamma_{[A}\chi^{\,a}\,A_{B]}^{\,a}\,.
\err
Note that the functions $\b(x^{11})$ and $\g(x^{11})$ are conceivably
related to auxiliary fields. 

Of particular interest are Bianchi identities 
satisfied by $G_{IJKL}$ defined by (\ref{gdef}).
The contribution to $dG$ with five ten-dimensional
indices vanishes. Thus, $5\der_{[A}G_{BCDE]}=0$.  But the components of
$dG$ with one eleven index become nontrivial.  We find
\bq 5\der_{[11}G_{ABCD]} =\bpl \gamma'(x^{11})-36\b(x^{11})\bpr
    F_{[AB}^aF_{CD]}^a \,.
\label{bi1}\eq
This expression is crucial for the analysis which follows.

In the following we restrict our analysis to the coupling of
the super Yang-Mills theory propagating on the fixed-hyperplane
at $x^{11}=0$.  A complementary analysis applies to the other
fixed hyperplane at $x^{11}=\pi$.  We omit a discussion of the
complementary coupling for reasons of economy.

\subsection{Close Algebra on Fermions (Construct Invariant Action)}
As described above, since we are working with on-shell
representations, the superalgebra closes
on fermions only if one invokes equations of motion
derived from an invariant action.  Therefore, we may not 
straightforwardly employ the techniques used in the previous subsection.
But it is sufficient to determine transformation rules 
which leave invariant an action.  If the action is
invariant under the transformation rules, then the rules must close
up to equations of motion.  One can force closure of the action
by first including the general modifications 
to the transformation rules obtained in the previous subsection, 
and then by adding suitable interaction
terms to the combined action.  Demanding invariance of the action 
under supersymmetry further constrains the modifications 
involving the functions $\b(x^{11})$ and $\g(x^{11})$ 
introduced above, but not completely.  All remaining
ambiguities are then removed by demanding invariance as well under
local Yang-Mills transformations and local Lorentz transformations.

The logical first step in modifying the action 
is to add a coupling between the gravitino field
and the supercurrent obtained by varying the
Yang-Mills action (\ref{action10}). 
This ``Noether" term is given by
\bq S_{\rm Noether}=\l^{-2}\,\int_{M^{10}}\,\bpl
    -\ft14e^{(10)}\bar{\psi}_A\Gamma^{BC}\Gamma^A\chi^{\,a}\,
    F_{BC}^{\,a}\,\bpr \,.
\eq
As described in \cite{hw2}, adding this term does not yet yield 
invariance.  Assorted further interactions are also needed.
Crucial is a necessary modification to the Bianchi identity
for $G_{IJKL}$, found to be
\bq 5\der_{[11}G_{ABCD]}=\mp\,3\sqrt{2}\,\frac{\k^2}{\l^2}\d(x^{11})\,
    F_{[AB}^aF_{CD]}^a \,,
\label{req}\eq
where $\d(x^{11})$ is a delta function, and
where the $\mp$ depends on whether $\Gamma_{11}\psi_A=\pm\psi_A$ at
the orbifold fixed-point.  The cancellation which results 
with the modification (\ref{req}) when verifying invariance of the action
requires an integration by parts.  
In the orbifold picture, which we 
use throughout this paper, there is no boundary, and therefore
the integration by parts does not involve a nontrivial boundary 
contribution.  The delta function in (\ref{req}) restricts the
required new coupling to the fixed hyperplane at $x^{11}=0$, where
the Yang-Mills fields are constrained to propagate.

It is reassuring that the form of this new coupling is consistent
with the result obtained in the previous subsection based on
a study of the gauge algebra.  
Now, we can reconcile this new result with our previous results to
obtain further constraints on the functions
$\gamma(x^{11})$ and $\b(x^{11})$.  Comparing     
(\ref{req}) with (\ref{bi1}) we determine that
\bq  \gamma'(x^{11})-36\b(x^{11})    
     =\mp 3\sqrt{2}\,\frac{\k^2}{\l^2}\d(x^{11}) \,.
\label{cond}\eq
A minimal solution to this equation, which restricts
all nontrivial features of the functions $\b(x^{11})$ and $\g(x^{11})$
to the fixed hyperplane at $x^{11}=0$ is the following,
\brr \gamma(x^{11}) &=& b\,\frac{\k^2}{\l^2}\,\t(x^{11})
     \nonumber\\
     \b(x^{11}) &=&
     \ft{1}{36}(2b\pm3\sqrt{2})\,\frac{\k^2}{\l^2}\,\d(x^{11}) \,,
\label{fns}\err
where $\t(x^{11})$ is the Heavyside (or step) function, which has the
property that $\t'(x^{11})=2\d(x^{11})$,   
and $b$ is a real dimensionless constant which cancels in (\ref{cond}).
\footnote{We could also add an arbitrary
smooth function $36 f(x^{11})$ to $\g(x^{11})$.
If we simultaneously added $f'(x^{11})$ to $\b(x^{11})$, 
this extra function would drop out of (\ref{cond}).  But we ignore such a possibility in this analysis.  However, it has been suggested that it may be fruitful to consider Yang-Mills fields not
rigidly constrained to the fixed hyperplanes, but rather 
somehow inhabiting a boundary layer.  In such a scenario one could envision the
delta function smearing out.  In that case, such a modification could
prove useful.}  We strongly emphasise that 
{\it the constant $b$ is not fixed by supersymmetry}.
However, it {\it is} fixed by the requirements of gauge invariance 
and local Lorentz invariance.  

To enforce invariance of the action, it is also necessary to
add some higher fermi terms to the action, and also to further modify the
supersymmetry transformation rules for the gravitino $\psi_I$ and
the gaugino $\chi^{\,a}$.  These modifications are all higher-fermi.
They are straightforward to compute, and they do not pose any 
obstruction to the results which we have already derived. 
Most significantly they do not pose any additional restrictions
on the value of the constant $b$. 
We therefore suppress these extra modifications in this paper.

It is useful to summarize the results which we have derived so far.
First, the four-form field strength must have the
following form.
\brr G_{ABCD} &=& 24\der_{[A}C_{BCD]}
     +b\,\frac{\k^2}{\l^2}\t(x^{11})F_{[AB}^aF_{CD]}^a 
     \nonumber\\[.1in]
     G_{(11)ABC} &=& 24\der_{[11}C_{ABC]}
     +\ft16(2b\pm 3\sqrt{2})\frac{\k^2}{\l^2}\d(x^{11})\w_{ABC} \,.
\label{modg}\err
Second, supersymmetry transformations of the 
three-form should include the following modifications to the
rule exhibited in (\ref{rules11}),
\brr \d_Q^{\,'}C_{ABC} &=&
     -\ft{1}{12}\,b\,\frac{\k^2}{\l^2}\,\t(x^{11})\,
     \bar{\e}\Gamma_{[A}\chi^{\,a}\,F_{BC]}^{\,a} 
     \nonumber\\[.1in]
     \d_Q^{\,'}C_{AB(11)} &=&
     \ft{1}{36}(2b\pm 3\sqrt{2})\,\frac{\k^2}{\l^2}\,\d(x^{11})\,
     \bar{\e}\Gamma_{[A}\chi^{\,a}A_{B]}^{\,a} \,.
\label{susymod}\err
Third, and most importantly, 
the Yang-Mills gauge transformation
should act on the mixed components of the three-form as follows,
\bq \d_\t C_{AB(11)}=\ft{1}{36}(2b\pm 3\sqrt{2})\frac{\k^2}{\l^2}
    \d(x^{11})\,\t^a\,F_{AB}^a \,.
\label{modc}\eq
Aside from the higher fermi terms which we have suppressed, the results
(\ref{modg}), (\ref{susymod}) and (\ref{modc}) represent the general
solution to the constraints obtained from implementing supersymmetry. 
There still remains the unfixed constant $b$.  
This is determined by imposing gauge invariance.
We consider this issue in the following subsection.

Since the step function $\t(x^{11})$ is discontinuous at $x^{11}=0$, the precise
boundary value of $G_{ABCD}$, as defined in (\ref{modg}),
is not well-defined.  However,
the following product does have a well-defined behavior,
\bq G_{[ABCD}G_{CDEF]}\,|=
    b^2\,\frac{\k^4}{\l^4}\,F_{[AB}^{\,a}\,F_{CD}^{\,a}\,
    F_{EF}^{\,b}\,F_{GH]}^{\,b} \,.
\label{bound}\eq
This relation is necessary for proving the gauge invariance 
of the theory.  We examine this issue presently.

\subsection{Implement Gauge Invariance}
Above, we considered the possibility that the three-form
$C_{IJK}$ may transform nontrivially 
under a Yang-Mills gauge transformation.
We therefore draw special attention to the following
term in the action,
\bq W=-\frac{\sqrt{2}}{\k^2}\,\int_{M^{11}}\, 
    C\wedge G\wedge G \,.
\eq
Under a Yang-Mills variation, this term may also transform 
nontrivially due to the nonvanishing of $\d_\t C_{AB(11)}$.  
In the classical case, such behavior would obstruct
gauge invariance and spoil the consistency of the theory.
In the quantum case, on the other hand, this behavior might be
welcome, as $W$ may provide a counterterm for
gauge anomalies.  This possibility was introduced in \cite{hw2}.
Using the transformation rule 
(\ref{modc}) as well as the boundary condition 
(\ref{bound}) it follows that $W$ transforms as follows,
\bq \d_\t W = -(2b\pm3\sqrt{2})\,b^2\,
    \frac{\sqrt{2}}{12}\,
    \frac{\k^4}{\l^6}\,
    \int_{M^{10}}\,\t^a F^a\wedge F^b\wedge F^b\wedge F^c\wedge F^c \,.
\label{anom}\eq
In the classical theory,
we can retain gauge invariance only if this term vanishes.
Therefore, in the classical theory, gauge invariance requires the
following condition,
\bq (2b\pm3\sqrt{2})\,b^2=0 \,.
\eq
There are exactly two solutions to this condition, $b=0$ or 
$b=\mp 3/\sqrt{2}$.  From the expressions in the previous
subsection, particularly (\ref{modg}),
we see that the first of these choices, $b=0$,  amounts to concentrating all
new features into components of $G_{IJKL}$ with
mixed indices.  The second choice, $b=\mp 3/\sqrt{2}$ amounts to 
concentrating all 
new features into the complementary components of $G_{IJKL}$,
or those which have purely ten-dimensional indices.  
From equation (\ref{modc})
we also see that the first choice, $b=0$, requires a nontrivial
Yang-Mills transformation law for $C_{AB(11)}$, whereas the second choice,
$b=\mp 3/\sqrt{2}$ allows $C_{IJK}$ to remain completely inert under Yang-Mills
transformations.  

Does the theory make a unique choice between the
two possibilities $b=0$ or $b=\mp 3/\sqrt{2}$?  Substituting the expressions
from the previous subsection into the action formula, one finds that
the choice $b=0$ gives rise to factors of $\d(0)$ in the 
action, whereas the other choice $b=\mp 3/\sqrt{2}$ does not.  In the quantum
theory there exist conceivable ways to regulate these factors.
But in the classical theory there is no obvious way to make sense of them.
In order to describe a well-defined classical theory, it it apparent
that only the choice $b=\mp 3/\sqrt{2}$ is permissible.  

Note that we have described a coupling in which the tensor
field remains inert under the Yang-Mills transformation.  Note 
also the contrast
with the familiar ten-dimensional coupling necessary for the low-energy
description of the heterotic string.  In that case, it was impossible to find 
such a coupling; the two-form was required to transform even in 
the classical theory.  In the weakly-coupled heterotic string case, however,
there is no classical obstruction since one can omit the
counter terms if one also avoids loop effects.  In the case of M-theory,
we require the Green-Schwarz-like interaction even in the 
classical theory, but we find that we may nevertheless formulate a consistent
classical theory which leaves the three-form inert under Yang-Mills
transformations.

\section{The Quantum Limit}
When we include quantum modifications to the effective action,
new effects, attributable to loop diagrams, contribute to gauge,
gravitational and mixed anomalies.  Particularly, there exists an
anomalous contribution to Yang-Mills gauge transformations of the type 
$\int_{M^{10}}\,\t\,F^5$.  The group theoretic factors in this 
expression organize into the form of (\ref{anom}) precisely 
and uniquely for the choice of gauge group $E_8$.  
When we compute the relevant 
coefficient, and combine the result with 
(\ref{anom}), gauge anomaly cancellation is found to require
\bq (2b+3\sqrt{2})\,b^2=\nu \,,
\label{cubic}\eq
where we have specialized to the case
where $\Gamma_{11}\psi_A\,|=\psi_A\,|$.
In equation (\ref{cubic}), the real parameter $\nu$ arises from an anomalous 
contribution to the variation of the effective action.  The precise value
of $\nu$ can be computed using the techniques explained in \cite{zumino}.
We find
\bq \nu=\frac{\sqrt{2}}{2(4\pi)^5}\frac{\l^6}{\k^4}\,.
\eq
In the classical theory, described above, we do not involve
loop effects, and so $\nu=0$.  In the quantum theory $\nu\ne 0$,
and Yang-Mills invariance requires $b$ to be a real root of the
cubic equation (\ref{cubic}).

For any value of $\nu$ there is at least one such solution,
given by 
\brr b &=& 
     4^{-1/3}\,\bpl\,\nu-\sqrt{2}
     +\sqrt{\,(\nu-2\sqrt{2})\,\nu\,}\,\,\bpr^{1/3}
     \nonumber\\
     & & +4^{-1/3}\,\bpl\,\nu-\sqrt{2}
     -\sqrt{\,(\nu-2\sqrt{2})\,\nu\,}\,\,\bpr^{1/3}
     \nonumber\\
     & & -\frac{1}{\sqrt{2}} \,.
\label{soln}\err
This equation straightforwardly gives the required value of $b$ for any given
value of $\nu\ge 2\sqrt{2}$.  The equation is also valid in the regime
$0\le\nu<2\sqrt{2}$, but in this case more care must be used.  In this 
second case the first two terms of (\ref{soln}) are complex, but the imaginary
parts cancel. The remaining two roots of (\ref{cubic}) are 
not real and are therefore irrelevant to us (since $b$ is necessarily
real) when $\nu>2\sqrt{2}$.  However, in the regime $0\le\nu<2\sqrt{2}$,
all three roots of (\ref{cubic}) are real.  The locus of permitted
values of the pair $(\nu, b)$ are shown in the figure.

At two special points,
coresponding to the turning points evident in the figure,
the two complex roots of (\ref{cubic}) become real and degenerate.
The first of these is at the point $\nu=0$, which coincides with the 
classical theory, as 
described above.  In that case, in addition to the solution 
$b=-\ft32\sqrt{2}$, there is another
solution $b=0$.  This degeneracy is described above, and is
removed by the fact that the $b=0$ solution gives rise to inscrutible
factors of $\d(0)$ in the action which cannot be removed by a
regulator in the classical case.  In the quantum case, 
these $\d(0)$ factors are unavoidable,
since we cannot choose $b=-\ft32\sqrt{2}$, as this would not be 
compatible with (\ref{cubic}).
But in the quantum case this is not as problematic, since then we 
{\it expect} short distance curiosities, and we expect that a mechanism
exists to regulate these factors.  An important point, however, is
that the existence of ultraviolet problems, like the factors of
$\d(0)$, are entirely distinct from the issue of
anomaly cancellation.  All gauge, gravitational and mixed
anomalies should cancel irrespective of short-distance oddities,
such as the factors of $\d(0)$.

\begin{figure}[t]
\begin{center}
\includegraphics[width=60mm,angle=-90]{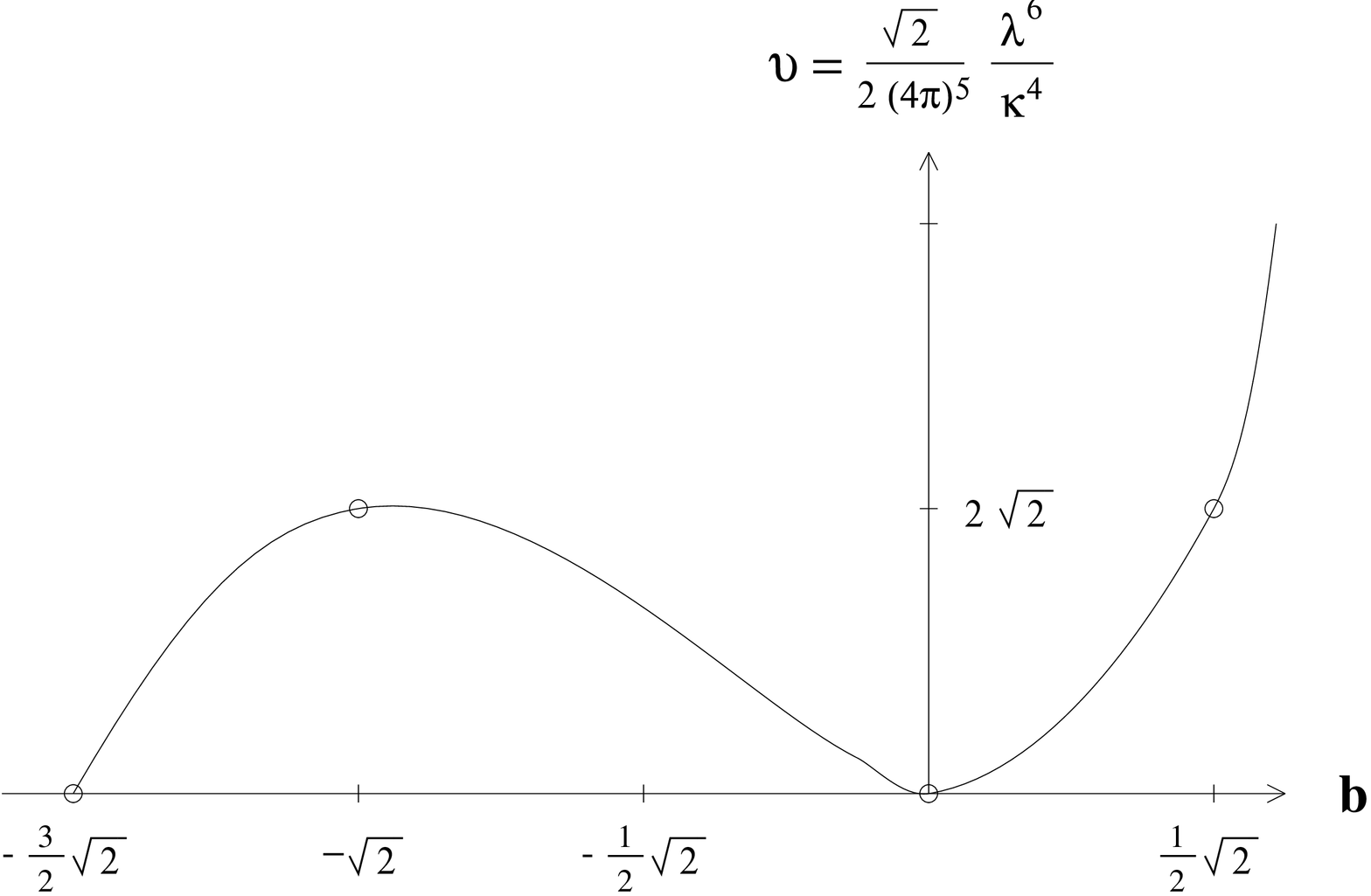}\\
\vspace{.2in}
\parbox{4in}{Figure: Curve indicating the value(s) of the parameter
$b$ permissible for gauge anomaly cancelation in M-theory, given the
values of the coupling constants $\l$ and $\k$, or vice-versa.} 
\end{center}
\end{figure}

There is one unique point in the quantum regime (where $\nu\ne 0$)
where the two comlex roots become real and
degenerate.  At this point $\nu=2\sqrt{2}$, and one root 
tells us that $b=\ft12\sqrt{2}$.
The additional degenerate roots tell us that we could also take
$b=-\sqrt{2}$.  This constitutes the second special point mentioned above.
As yet we have no reason to believe that the theory selects this point.

The cancelation of all anomalies requires a further modification
to the Bianchi identity for $G_{IJKL}$ discussed above, so that it
involves factors of $R_{[IJ}R_{KL]}$ and/or the Lorenz Chern-Simons
form.  The probability that the gravitational anomalies can be
eliminated depends on nontrivial factorization
properties required on the precise structure of the anomalies. 
Ho{\v r}ava and Witten have demonstrated in \cite{hw1}
and \cite{hw2} that these properties are indeed satisfied.
The mechanism requires more Green-Schwarz-like
counterterms, crudely of the form $C\wedge R^4$.  Since the
three-form has the Yang-Mills transformation property 
determined above by the elimination of pure gauge anomalies, and since
this rule involves the yet-undetermined parameter $b$,
it is clear that the value of $b$ is selected only
from a knowledge of the precise coefficients describing the
gravitational or mixed anomalies.  As described in \cite{hw2},
it is not a simple matter to determine these numbers.  
We will discuss this issue in detail in a forthcoming paper.  

\section{Phenomenological Ramifications and Conclusions}
The essential modification which enables the coupling between
eleven-dimensional supergravity and fixed-point ten-dimensional 
super Yang-Mills theory is exhibited in
(\ref{modg}).  From that equation we see that the fundamental
parameter governing the coupling is $\k^2/\l^2$.  
However, we have seen that
consistency requires that $\l^2\sim \k^{4/3}$, with a precise numerical
coefficient determined by the requirements described above.
Thus, the fundamental expansion parameter is $\k^{2/3}$.
Without presenting details of higher-fermi modifications, we
have described effects occuring at lowest-order in this
parameter which are required by supersymmetry and by anomaly 
cancelation.  These are summarized by equation (\ref{modg}).
In the quantum theory, the parameter $b$ is a real root of the 
cubic equation  (\ref{cubic}).  Since the value of $\nu$ 
depends on the ratio $\l^6/k^4$,
further constraints are required to completely fix its value.
These are provided by gravitational or mixed anomaly cancelation.

To leading order in $\k^{2/3}$ all of the purely bosonic contributions
to the action are obtained by substituting (\ref{modg}) into
(\ref{action11}).  In the classical case, there are no terms
involving $\d(0)$ because the choice $b=\mp3/\sqrt{2}$ removes
them.  In that case,
terms involving $\t(x^{11})$ appear either squared,
and are therefore relatively well defined, 
or vanish due to the fixed-point conditions on the various fields.
It is apparent that considerations similar to those
already discussed will apply at higher orders
in $\k^{2/3}$.  Thus, in the classical limit we only retain modifications
which are relatively well defined 
(ie: $\t(x^{11})$ appears raised only to even powers.). 
The complete justification of this statement
of course necessitates that one computes these higher-order modifications.

To date, all attempts at describing M-theory phenomenology
have had an endemic problem with factors of $\d(0)$.  This issue
is dealt with at levels of rigor ranging from a (low) of sweeping
the issue completely under the carpet to a (high) of 
demonstraing some purely formal cancelations or by
systematically classifying these factors as an effect occurring at
higher order in the expansion parameter $\k^{2/3}$.  In \cite{peskin},
this isssue was addressed in an analogous situation in five-dimensions,
where the factors of $\d(0)$ were shown to derive from the
elimination of an auxiliary field.   In the quantum theory, the proper 
regulation of the terms formally proportional to $\d(0)$ has yet
to be fully explained.  

In the current analysis, we find that we can at least move the issue of
the delta functions one rung down the ladder of relevance, by suggesting
a classical limit completely devoid of these factors. 
This fact opens the door to a more ``honest" phenomenology.

There is enough information about the low-energy behavior
of M-theory to make concrete predictions about nature.
At the same time, it should be possible to
discuss a classical limit. That is, we should be able to take
$\hbar\rightarrow 0$, and obtain meaningful leading-order results
from these predictions, without having to rationalize the neglect of
otherwise inscrutible terms such as those containing formal 
factors of $\d(0)$. In this paper we have described a coherent 
way of understanding this classical limit.

At the same time we have discussed a new ingredient necessary for
implementing proper anomaly cancellation in these theories, 
We have argued that the web of constraints which include all gauge, gravitational and mixed anomalies is more rich than 
previously understood.
We argue that the new parameter, called $b$ in this paper, is necessary
to ensure that the full complement of consistency requirements
in the theory not be overconstrained.
A detailed analysis of precise anomaly coefficients is necessary to 
reconcile this issue.  
This will be discussed in a forthcoming paper.

\vspace{.3in}
\noindent
{\Large{\bf Acknowledgement}}\\[.2in]
I acknowledge useful discussions with Dieter L{\"u}st, 
Christian Prietchopf, and Andr{\'e} Lukas.    
I especially thank Burt Ovrut for his hospitality at
the University of Pennsylvania, where some of this work was accomplished.


\begin{thebibliography}{99}
%
\bibitem{wittenstring}
{\it String Theory Dynamics in Various Dimensions}\\
E.Witten, hep-th/9503124
%
\bibitem{hw1}
{\it Heterotic and Type I String Dynamics from Eleven Dimensions}\\
P.Ho{\v r}ava and E.Witten, hep-th/9510209
%
\bibitem{hw2}
{\it Eleven-Dimensional Supergravity on a Manifold with Boundary}\\
P.Ho{\v r}ava and E.Witten, hep-th/9603142
%
\bibitem{dienes}
{\it String Theory and the Path to Unification: A Review of Recent
Developments}\\
K.R.Dienes, Phys. Rep. 287 (1997) 447, hep-th/9602045
%
\bibitem{dm}
{\it On the Strongly Coupled Heterotic String}\\
E.Dudas and J.Mourad, Phys. Lett. 400B (1997) 71, 
hep-th/9701048
%
\bibitem{lu}
{\it Remarks on M-theory Coupling Constants and M-Brane Tension
Quantizations}\\
J.X.Lu, hep-th/9711014
%
\bibitem{ns}
{\it String-Unification, Universal One-Loop Corrections
and Strongly Coupled Heterotic String Theory}\\
H.P.Nilles and S.Stieberger, 
hep-th/9702110
%
\bibitem{low}
{\it On the Four-Dimensional Effective Action of Strongly Coupled
Heterotic String Theory}\\
A.Lukas, B.A.Ovrut and D.Waldram, 
hep-th/9710208
%
\bibitem{dg}
{\it Four-Dimensional M-theory and supersymmetry breaking}\\
E.Dudas and C.Grojean,
hep-th/9704177
%
\bibitem{noy}
{\it Supersymmetry Breaking and Soft Terms in M-Theory}\\
H.P.Nilles, M.Olechowski and M.Yamaguchi,
hep-th/9707143
%
\bibitem{aq}
{\it Supersymmetry breaking in M-theory}\\
I.Antoniadis and M.Quir{\'o}s,
hep-th/9709023
%
\bibitem{peskin}
{\it Transmission of Supersymmetry Breaking from a 4-dimensional Boundary}\\
E.A.Mirabelli and M.E.Peskin, hep-th/9712214.
%
\bibitem{cjs}
{\it Supergravity Theory in 11 Dimensions}\\
E.Cremmer, B.Julia and J.Scherk, 
Phys. Lett. 76B (1978), 409-412
%
\bibitem{chamseddine}
{\it $N=4$ Supergravity Coupled to $N=4$ Matter}\\
A.Chamseddine, Nucl. Phys. {\bf B}185 (1981) 403.
%
\bibitem{bdrdwvn}
{\it Ten-Dimensional Maxwell-Einstein Supergravity, its Currents
and the Issue of its Auxiliary Fields}\\
E.Bergshoeff, M.de Roo, B.de Wit and P.van Nieuwenhuizen,
Nucl. Phys. {\bf B}195 (1982) 97.
%
\bibitem{cm}
{\it Unification of Yang-Mills Theory and Supergravity in Ten Dimensions}\\
G.F.Chapline and N.S.Manton,
Phys. Lett. 120B (1983) 105-109
%
\bibitem{gs}
{\it Anomaly Cancellations in Supersymmetric $D=10$ Gauge Theory
and Superstring Theory} \\
M.B.Green and J.H.Schwarz, 
Phys. Lett. 149B (1984), 117-122.
%
\bibitem{zumino}
{\it Chiral Anomalies, Higher Dimensions, and Differential Geometry}\\
B.Zumino, Y.S.Wu and A.Zee, Nucl. Phys. {\bf B}239 (1984), 477-507.
%

\end{thebibliography}
\end{document}